\documentclass[11pt]{article}
\usepackage[margin=1.25in]{geometry}
\usepackage[usenames,dvipsnames]{color}
\usepackage{url}
\usepackage[colorlinks = true,
            linkcolor = blue,
            urlcolor  = blue,
            citecolor = blue,
            anchorcolor = blue]{hyperref}

\usepackage{graphicx,amsmath,amsfonts,amscd,amsthm,amssymb,pstricks}
\usepackage{color}
\usepackage{dcolumn}
\usepackage{bm}
\usepackage{longtable}
\usepackage{mathrsfs}
\usepackage{textcomp}
\usepackage{esvect}
\usepackage{float}
\usepackage[USenglish,british]{babel}
\usepackage{environ}
\usepackage{xifthen}
\usepackage{xargs}		
\usepackage{hyperref}
\usepackage[separate-uncertainty = true]{siunitx}
\DeclareSIUnit\barn{b}
\usepackage{physics,mathtools}
\usepackage{multirow,tablefootnote}
\usepackage{comment}
\usepackage{enumerate}
\usepackage{appendix}
\usepackage{tikz}	
\usetikzlibrary{arrows, calc, matrix, patterns, decorations.markings, shapes, decorations.pathmorphing, shadows.blur}
\usepackage[utf8]{inputenc}
\usepackage{enumitem}

\newcommand\pubnumber{}
\newcommand\pubdate{\today}

\def\Title#1{\begin{center} {\LARGE #1 } \end{center}}
\def\Author#1{\begin{center}{ \sc #1} \end{center}}
\def\Address#1{\begin{center}{ \it #1} \end{center}}

\newcommand\pubblock{\rightline{\begin{tabular}{l} \pubnumber\\
         \pubdate \end{tabular}}}
\newenvironment{Abstract}{\begin{quotation} \begin{center}
                       ABSTRACT
     \end{center}\bigskip  }{\end{quotation}}

\newcommand\snowmass{\begin{center}\rule[-0.2in]{\hsize}{0.01in}\\\rule{\hsize}{0.01in}\\
\vskip 0.1in Submitted to the  Proceedings of the US Community Study\\ 
on the Future of Particle Physics (Snowmass 2021)\\ 
\rule{\hsize}{0.01in}\\\rule[+0.2in]{\hsize}{0.01in} \end{center}}

\begin{document}

\pubblock

\Title{Future Circular Hadron Collider FCC--hh: Overview and Status}

\bigskip 

\Author{M. Benedikt$\,^1$, A. Chance$\,^3$, B. Dalena$\,^3$, D. Denisov$\,^2$, M. Giovannozzi$\,^1$, J.~Gutleber$\,^1$, R.~Losito$\,^1$, M. Mangano$\,^1$, T. Raubenheimer$\,^5$, W. Riegler$\,^1$, V.~Shiltsev$\,^4$, D. Schulte$\,^1$, D. Tommasini$\,^1$, F. Zimmermann$\,^1$}

\medskip

\Address{$^1$ CERN, Esplanade\ des Particules 1, 1211 Geneva 23, Switzerland\\
$^2$ Brookhaven National Laboratory, Nuclear and Particle Physics, Upton, NY, USA\\
$^3$ Commissariat \`a l'\'energie atomique et aux \'energies alternatives - Institut de Recherche sur le lois Fondamentales de l'Univers Saclay (CEA/Irfu Saclay), Gif-sur-Yvette, France \\
$^4$ Fermi National Accelerator Laboratory (FNAL), Batavia, IL, USA \\
$^5$ Stanford National Accelerator Center (SLAC), Menlo Park, CA, USA}

\medskip

 \begin{Abstract}
\noindent The Future Circular Collider (FCC) study was launched as a world-wide international collaboration hosted by CERN. Its goal is to push the field to the next energy frontier beyond LHC, increasing by an order of magnitude the mass of particles that could be directly produced, and decreasing by an order of magnitude the subatomic distances to be studied. The FCC study covers two accelerators, namely, an energy-frontier hadron collider (FCC--hh) and a highest luminosity, high-energy lepton collider (FCC--ee). Both rings are hosted in the same 100 km tunnel infrastructure, replicating the CERN strategy for LEP and LHC, i.e.\ developing a lepton and a hadron ring sharing the same tunnel. This paper is devoted to the FCC--hh and summarizes the key features of the FCC--hh accelerator design, performance reach, and underlying technologies. The material presented in this paper builds on the conceptual design report published in 2019, and extends it, including also the progress made and the results achieved since then.
\end{Abstract}

\snowmass

\def\thefootnote{\fnsymbol{footnote}}
\setcounter{footnote}{0}
\section{Design overview}

\subsection{Status}

The design of the FCC--hh collider has been presented in a Conceptual Design Report (CDR)~\cite{FCC-hhCDR}, which describes the baseline configuration of the ring (see Section~\ref{sec:intro} and following for a brief review of the baseline design and of the recent developments). Note that the discussion presented in the rest of this paper is essentially based on the material collected in the CDR. 

Since the publication of the CDR, substantial progress has been made, in particular in the domain of ring placement and layout, and the main results are summarized in Section~\ref{sec:progress}.

\subsection{Performance matrix}

\subsubsection{Attainable energy}

The target energy of \SI{100}{TeV} fully relies on the successful development of \SI{16}{T},  superconducting magnets, and any failure to meet the target magnetic field will necessarily impact the final energy of the collider. To mitigate the risk linked to this challenging and new technology, R\&D efforts are needed and accurately detailed in~\cite{FCC-hhCDR}. 
In this respect, the experience from HL--LHC will be important.

\subsubsection{Attainable luminosity and luminosity integrals}

Possible limiting factors for the collider luminosity seem more linked to luminosity integrals rather than attainable luminosity. 

In the case of the LHC, the attainable luminosity has surpassed the nominal one thanks to several elements. Higher-brightness beams delivered by the injectors boosted the luminosity, while in the LHC ring the excellent optics control, which includes measurement and correction, together with an optimal use of the available beam aperture, thanks also to the use of tighter collimators settings~\cite{FusterLHCapertureEPJP}, provided the final touch. It is worth stressing that in the LHC, $\beta^\ast=\SI{30}{cm}$, corresponding to the nominal FCC--hh value, has already been achieved and surpassed. On the downside, the larger number of quadrupole magnets in the straight sections of the FCC--hh might challenge the correction algorithms devised so far for the LHC, and new approaches should be explored. Furthermore, the actual operational performance with crab cavities is still unknown, but the HL--LHC will provide an ideal test-bed for getting ready in view of the FCC--hh.

In this respect, attaining the FCC--hh target integrated luminosity might be more challenging for several reasons. The injector chain will increase in terms of the number of accelerator rings; the number of magnets (and of active elements in general) in the FCC--hh lattice will also increase with respect to the LHC or HL--LHC; repairing activities will be challenging, also taking into account the distances to be covered to access the faulty hardware and the large number of components. All these considerations suggest that operational efficiency might be at risk, and that appropriate mitigation measures should be considered (e.g. repairing activities carried out by robots).

\subsubsection{Injector and driver systems}

The baseline option for the FCC--hh ring is to use the LHC injector chain and the LHC as pre-injector. An alternative consists of replacing the LHC in its role of pre-injector with a superconducting ring to be installed in the SPS tunnel. The LHC injector chain is, with no doubt, a key element in the success of the LHC. The increase in its complexity, with the addition of the LHC, will potentially impact on its reliability. Furthermore, the various accelerators in the injector chain will have rather different ages, with the Proton Synchrotron being the oldest ring (it was commissioned in November 1959). This might have impact on the overall performance and should be properly addressed, e.g. with an appropriate long-term maintenance programme.  

\subsubsection{Facility scale}

\begin{figure}[htb]
\begin{center}
\includegraphics[trim=2truemm 0truemm 2truemm 0truemm,width=0.60\hsize,clip=]{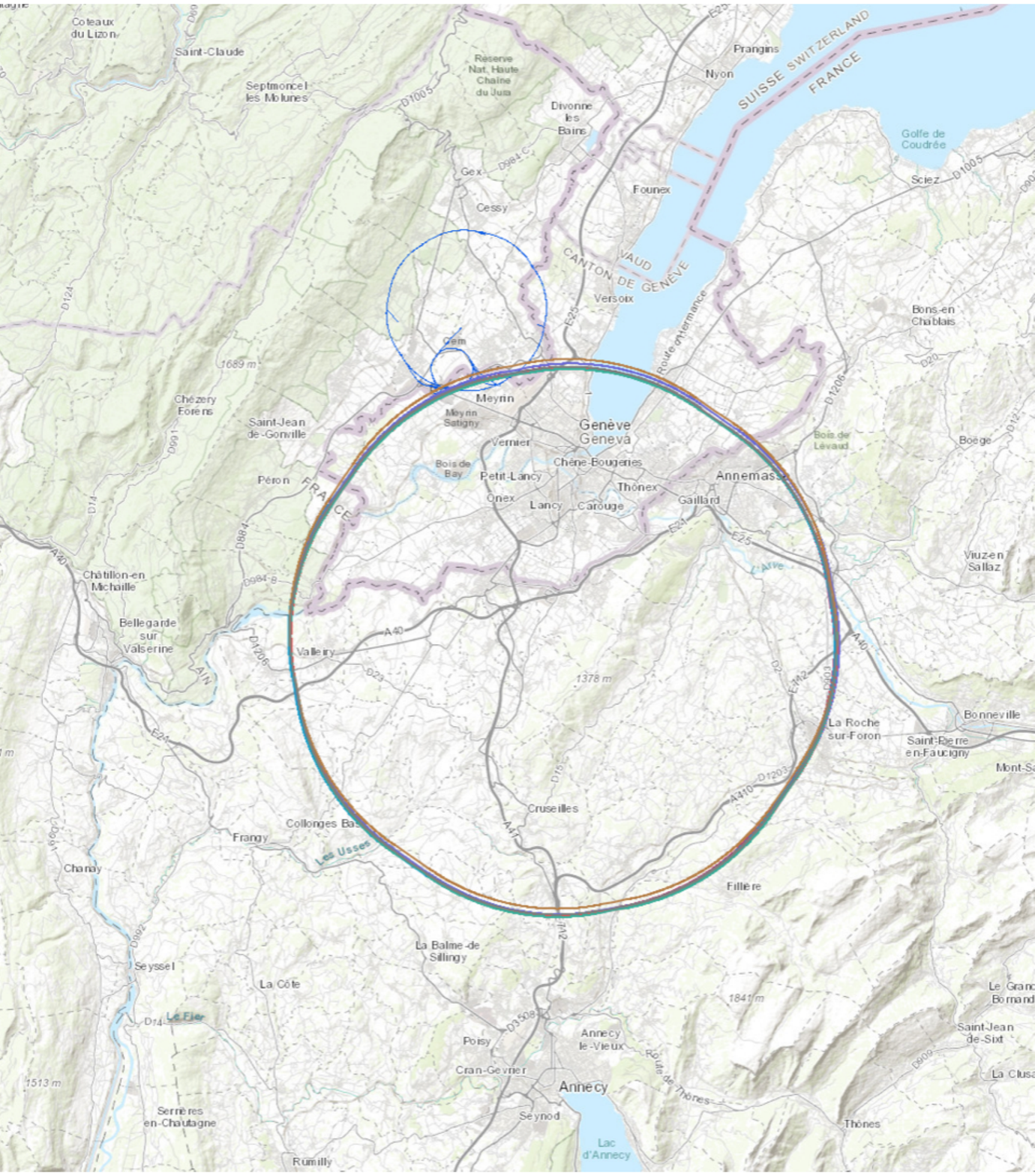}
\end{center}
\caption{Picture of some of the FCC-hh implementation variants under study, including the LHC and the SPS accelerators.}
\label{fig:placement}
\end{figure}

Figure~\ref{fig:placement} shows some of the FCC-hh implementation variants under study, including the LHC and the SPS accelerators. The large scale of the FCC--hh ring and the related infrastructure, implies a certain number of risk factors stemming from the civil engineering activities. The tunneling activities (for the ring tunnel as well as the ancillary tunnels) are comparable to those of the recently completed Gotthard Base tunnel (total of about \SI{152}{km}, including two \SI{57}{km} tunnel tubes) in Switzerland. Nevertheless, the handling of excavation materials might pose problems. In this respect, mitigation measures have been put in place in terms of R\&D for finding efficient treatment and use of these materials. As far as the infrastructure on the surface is concerned, possible difficulties might arise because of the regional and national frameworks in the two Host States that regulate the acceptance of an infrastructure development project plan. In this respect, close contacts have been established with national regulatory bodies to mitigate this risk.

\subsubsection{Power requirements}

The FCC--hh collider complex is expected to require about \SI{580}{MW} of electrical power,  which could be reduced to about \SI{550}{MW} with further optimization. Of these \SI{550}{MW}, about \SI{70}{MW} are needed for the injector complex, \SI{70}{MW} for cooling, ventilation, and general services. A further \SI{40}{MW} are consumed by the four physics detectors, and \SI{20}{MW} are allocated to the data centers for the four experiments. Among all the subsystems, the highest demand comes from the FCC--hh cryogenics, which requires about \SI{276}{MW} (about \SI{250}{MW} after further optimization, to be compared with about \SI{40}{MW} for the existing cryoplants of the LHC, with a three times shorter ring circumference), roughly half of which is needed to extract the $\sim \SI{5}{MW}$ of FCC--hh synchrotron radiation heat load from inside the cold arcs. These power requirements  were obtained thanks to a careful optimization of the FCC--hh components, and, in particular, by an optimized beam-screen temperature, energy-efficient designs, and the use of new energy-saving technologies. Note that losses in the power transmission corresponding to about 5\%-7\% of the peak power should be added to estimate the needed grid power. 

In addition to the successful efforts in optimizing the power consumption of the FCC--hh, attempts to further decrease the power needed are planned. These studies will be essential to improve the energy efficiency of the collider and thus enhance the public acceptance of this large-scale facility. The three-pronged strategy, put in place already since the conceptual design study phase, envisages a reduction of energy consumption, increase of efficiency of energy use, and the recovery and reuse of energy for other purposes. This strategy will further be pursued in the next phase of the FCC--hh study.  

\subsection{Challenges}

Although the FCC--hh clearly poses a number of possible obstacles in several areas (beam physics and technology), it builds on the experience of previous operational colliders, such as LHC and HL--LHC, which ensures the possibility of developing sound mitigation measures for the various challenges. As an example, it is worth mentioning that the machine design heavily relies on that of the LHC and HL--LHC, which instills confidence in the projected performance.

The unprecedented beam energy of \SI{8.3}{GJ} represents a challenge for all systems devoted to protecting the hardware integrity of the FCC--hh ring, such as the collimation and dump systems. Such a challenge translates into beam dynamics challenges, e.g. the optics design of for the straight sections housing collimation and dump systems, which should satisfy multiple constraints, such as phase-advance relations, beam aperture constraints, and beam impedance, just to mention a few. The requirements also bring technological challenges in several areas, e.g.in terms of materials selected for the collimators jaws, and beam dumps, but also for the hardware related to the kickers that are used to dump the beams and to dilute them before interacting with the dump material. 

The field quality of the main magnets at injection energy is also an aspect that deserves particular care, as an insufficient field quality might lead primarily to beam loss and possible also emittance growth, with a direct impact on machine performance. 

The technology upon which the FCC--hh design relies is that of high-field Nb$_3$Sn superconducting magnets. Multiple challenges can be identified, linked to different aspects of this hardware. For instance, one challenge is the development of the Nb$_3$Sn wire to sustain the high critical currents needed to achieve the \SI{16}{T} magnetic field. Such a goal should be achieved with the constraint that the wire be economically affordable, given the large-scale production of magnets needed for FCC--hh. An appropriate magnet design is another challenge, as this goal should be achieved by fulfilling several criteria, such as the minimization of the amount of superconductor and the field quality at injection energy. The complexity of this novel hadron collider is such that several other technologies are a key to implementing the FCC--hh. The most important ones are an efficient and cost-effective cryogenic refrigeration, superconducting septum magnets, and solid state generators. The best candidate for better (with respect to the LHC and HL--LHC choice) cryogenic refrigeration is based on a mixture of neon and helium. Superconducting septum magnets are essential for a compact and efficient design of the beam-dump system. Modular, scalable, fast, and affordable high-power switching systems are key components  of beam transfer systems. Solid-state devices, currently not commercially available, offer high-performance capabilities, which are needed for efficient FCC--hh operation.   

These technologies, which are connected with an overall increase of the operation efficiency of accelerator systems, naturally lead to the consideration of environmental aspects linked to the FCC--hh. Such a large-scale facility has an unavoidable impact on the environment due to the civil engineering works, radioactive waste, and energy efficiency. Concerning the first two aspects, CERN has a long experience due to the LEP/LHC experience. Although the FCC--hh scale exceeds by far the LEP/LHC scale, since the beginning of the studies, the respect of the environment and the minimization of the impact on it has been the main guideline. This criterion is applied not only to the underground infrastructure, but also to the surface infrastructure, given its direct societal impact. The radioactive waste management is a delicate aspect, but all means have been put in place to integrate it since the beginning in the global implementation project. Concerning energy efficiency, it is clear that this aspect is new and high in the societal opinion; for this reason several options have been studied and are actively pursued to provide more efficient energy consumption, e.g. via a new cryogenic system, as well as to recover, whenever possible, energy, which is the case of the waste heat recovery.

\section{Technology requirements} 

The technological choices presented in the FCC--hh CDR represent feasible options for the implementation of the hadron collider. The time needed to move from the CDR stage to a TDR stage allows for carrying out R\&D studies to pursue the detailed feasibility assessment of the various technological items that are comprised in the FCC--hh baseline. A set of so-called strategic R\&D topics have been identified, which are essential prerequisites for the preparation of a sound technical design.  

It is clear that in addition to the several technological challenges, a crucial aspect to consider and to assess carefully is the large-scale production of the \SI{16}{T} magnets. 

It is worth stressing that a detailed analysis of the possibility to establish partnerships has been carried out, and a series of universities and research institutes have been identified as possible partners. Furthermore, whenever possible and appropriate, industrial partners have been also identified. 

The list of the strategic R\&D topics is as follows
\begin{itemize}
    \item 16 Tesla superconducting high-field dual aperture accelerator magnet.
    \item Cost-effective and high-performance Nb$_3$Sn superconducting wire at industrial scale.
    \item High-temperature superconductors. The integrated project time line allows for the exploration and development of high-temperature superconductor (HTS) magnet technology, and of possible hybrid magnets, enabling improved performance, i.e.\ higher fields, or higher temperature. HTS options might be more rewarding than Nb$_3$Sn  technology, as they might allow for higher fields, better performance, reduced cost, or higher operating temperature and for this last aspect, HTS could be game changers.
    \item Energy efficient, large-scale cryogenic refrigeration plants for temperatures down to \SI{40}{K}.
    \item Invar-based cryogenic distribution line.
    \item Superconducting septum magnet (to be merged with high-power switching elements).
    \item High-speed, high-power switching system for beam transfer elements.
    \item Decentralized, high-capacity energy storage and release.
    \item Advanced particle detector technologies.
    \item Efficient and cost-effective DC power distribution.
    \item Efficient treatment and use of excavation material.
\end{itemize}

\subsection{High-Field Magnet R\&D}

The primary technology of the future circular hadron collider, FCC--hh, is the high-field magnets, and both high-field dipoles and quadrupoles~\cite{FCC-hhCDR} are required, or, possibly, combined-function magnets~\cite{our_paper6}. 

For constructing the accelerator magnets of the present LHC, the Tevatron, RHIC, and HERA, wires based on Nb-Ti superconductor were used. However, Nb-Ti magnets are limited to maximum fields of about \SI{8}{T}, as being operated at the LHC. The HL--LHC will, for the first time in a collider, deploy some tens of dipole and quadrupole magnets with a peak field of \SI{11}{}--\SI{12}{T}, based on a new high-field magnet technology using a Nb$_3$Sn superconductor. 
The Nb$_3$Sn superconductor holds the promise to approximately double the magnetic field, from $\sim \SI{8}{T}$ at the LHC to \SI{16}{T} for the FCC--hh.  

Recently, several important milestones were accomplished in the development of high-field Nb$_3$Sn magnets. At CERN, a block-coil magnet, FRESCA2, with a \SI{100}{mm} bore, achieved a world-record field of \SI{14.6}{T} at \SI{1.9}{K}~\cite{Willering:2019hhu}. In the US, a Nb$_3$Sn cosine-theta accelerator dipole short-model demonstrator with \SI{60}{mm} aperture~\cite{zlobin-napac19}, reached a similar field, of \SI{14.5}{T} at \SI{1.9}{K}~\cite{US-MDP:2021weg}.  

Forces acting on the magnet coils greatly increase with the strength of the magnetic field, while, at the same time, most higher-field conductors, such as the brittle Nb$_3$Sn, tend to be more sensitive to pressure. Therefore, force management becomes a key element in the design of future high-field magnets.   

Beside the development of optimized magnet design concepts, such as canted cosine-theta dipoles~\cite{caspi2013canted}, higher field can be facilitated by a higher-quality conductor. 
A Nb$_3$Sn wire development programme was set up for the FCC~\cite{8629920}. For Nb–Ta–Zr alloys, it could be demonstrated that an internal oxidation of Zr leads to the refinement  of Nb$_3$Sn grains and, thereby, to an increase of the critical current density~\cite{buta}. The phase evolution of Nb$_3$Sn wire during heat treatment is equally under study, as part of the FCC conductor development programme in collaboration with TVEL, JASTEC, and KEK~\cite{9369038}.  

Advanced Nb$_3$Sn wires including Artificial Pinning Centers (APCs) developed by two separate teams (FNAL, Hyper Tech Research Inc., and Ohio State; and NHMFL, FAMU/FSU) achieved the target critical current density for FCC, of \SI{1500}{A \per mm^2} at 16 T~\cite{uswire1, uswire2}, which is 50\% higher than for the HL--LHC superconductor. The APCs decrease the magnetization heat during field ramps, improve the magnet field quality at injection, and reduce the probability of flux jumps~\cite{xu2014refinement}.

In addition to Nb$_3$Sn wires, also high-temperature superconductors (HTS) are of interest, since they might allow for higher fields, operation at higher temperature, and, ultimately, perhaps even lower cost. In this context, the FCC conductor programme has been exploring the potential of ReBCO (Rare-earth barium copper oxide) coated conductors (CCs). In particular, the critical surfaces for the current density, $J_\mathrm{c} (T, B, \theta)$, of coated conductors from six different manufacturers: American Superconductor Co.~(US), Bruker HTS GmbH~(Germany), Fujikura Ltd~(Japan), SuNAM Co. Ltd~(Korea), SuperOx ZAO~(Russia) and SuperPower Inc.~(US) have been studied~\cite{senatore2015}.

Outside the accelerator field, HTS magnet technology could play an important role for fusion research. A number of companies are developing HTS magnets in view of fusion applications. One of these companies is Commonwealth Fusion Systems, which, in partnership with MIT’s Plasma Science and Fusion center, is designing SPARC, a compact net fusion energy device~\cite{sparc}. The SPARC magnets are based on second generation ReBCO conductors. Recently, the SPARC team successfully demonstrated a coil with \SI{20}{T} field~\cite{mitnews}. An interesting view on HTS prospects is presented in a Snowmass 2020 Letter of Interest~\cite{snowmassmatias}, according to which the actual material and process costs of HTS tapes are a small fraction of their current commercial value, and that there is a historical link between manufactured volume and price~\cite{mikek}.

\section{Staging options and upgrades}

Considerations on possible staging options for the FCC--hh can be made on the basis of the experience of LHC and HL--LHC. Various types of energy upgrade, from a limited one (of about 7\%, based on the assumed engineering margins of the the various systems and in particular of the main dipoles) to a major one (of about 93\%, the so-called HE--LHC~\cite{HE-LHCCDR}, based on FCC--hh-type main dipoles to be installed in the LHC tunnel) have been considered, but no one has been retained as an efficient upgrade path. On the other hand, upgrade of the luminosity has been approved as the route to improve the LHC performance within the LHC Luminosity Upgrade Project~\cite{BejarAlonso:2749422}. It is worth noting that the LHC luminosity upgrade goal is achieved thanks to changes in the LHC ring, leading to a reduction of $\beta^\ast$, but also to the upgrade of the injectors chain to generate brighter beams and higher currents~\cite{Damerau:1976692}. 

We may, therefore, speculate that an energy upgrade is not a realistic option for FCC--hh, unless even higher-field magnets, e.g.~based on HTS, became available. 

Instead, a luminosity upgrade, following the two FCC--hh stages already foreseen, with an ideal delivery of about \SI{2}{} or \SI{8}{\femto \barn^{-1}} per day, respectively, could be considered an interesting option. However, further reducing $\beta^\ast$ (the nominal value in the second stage of FCC--hh is \SI{30}{cm}) does not much increase the integrated luminosity without a higher proton intensity.  Already, as designed, the FCC--hh machine is cycling for about half of the time (with fairly demanding assumptions on the ramp speed of the injectors, either a slightly modified LHC or a new superconducting SPS), and the protons are burnt off quickly in collision (see Fig.~4 in~\cite{PhysRevSTAB.18.101002} and the associated equations). 
Burning off the protons even more quickly cannot much raise the integrated luminosity, but will mostly increase the event pile up. On the other hand, the FCC--hh baseline only considers rather moderate intensities from the injector of $\sim 10^{11}$ protons per bunch and \SI{0.5}{A} beam current, which are at least a factor $\sim 2$ below the capabilities of the upgraded LIU/HL--LHC complex.

Brightness of the injected beam is not a critical issue for FCC--hh, since the radiation damping will anyhow shrink the beam in the collider. Maximum integrated luminosity could be attained by exploiting the maximum proton rate, bunch intensity, and beam current available from the CERN LHC complex. However, the beam current in the FCC--hh rings is limited due to the SR heat load and the associated cryogenics power requirements. With HTS magnets operating at an elevated temperature, these  cryogenics needs would be relaxed, and a higher beam current might become possible.

Another possible approach would be not to cycle the FCC--hh collider, but to run it at constant magnetic fields and approximately constant beam current, using a top-up injection scheme as was successfully implemented for the two B-factories, is in routine use at the present SuperKEKB, and forms a key ingredient of the future FCC--ee lepton collider. For the case of FCC--hh, top-up injection requires the installation of a fast ramping \SI{50}{TeV} full energy injector, which might become available thanks to advancing magnet technology. To facilitate the design, the beam current in the top-up injector could be restricted to e.g. 10\% of the collider beam current. Such a top-up injector could increase the integrated luminosity of the FCC--hh by a significant factor.

Lepton colliders utilize radiation damping to merge injected particles with the stored beam. If the radiation damping in FCC--hh proved too slow for this purpose, the merging of injected and stored beams could be accomplished by other methods, e.g., by injection into nonlinear resonance islands, which are then collapsed~\cite{PhysRevSTAB.10.034001,PhysRevSTAB.18.074001}, or, alternatively, by innovative damping of the injected beam, e.g.~through optical stochastic cooling or coherent electron cooling. So, in short, the use of HTS magnets allowing for higher beam current or the installation of a novel fast cycling full energy top-up injector would be two plausible paths to increase the integrated luminosity of FCC--hh by a significant factor.

It is also worth mentioning that heavy-ion collisions are part of the FCC  baseline, although they formally represent an extension with respect to the proton-proton programme. However, lepton-hadron collisions, the so-called FCC--eh, are not part of the baseline and would be an appealing upgrade. 

Other extensions of the FCC--hh scope could be collisions with isoscalar light ion beams~\cite{KRASNY2020103792}, the realization of a Gamma Factory~\cite{krasny2015gamma,Krasny:2690736,krasny2020gamma}, and becoming an ingredient of a high-energy muon collider~\cite{MAP,IMC,JINSTMC,Antonelli:2015nla,Zimmermann:2018wfu,Zimmermann-ipac22}.

\section{Synergies with other concepts and/or existing facilities}

Clearly, a natural synergy exists between FCC--ee and FCC--hh. Moreover, The FCC--hh can profit from the experience of LHC/HL--LHC in several aspects. The HL--LHC bases its luminosity increase upon the use of Nb$_3$Sn quadrupoles for the final focus. Hence, the experience gained in the design, prototype, construction, test, and operation of the new triplets will be essential for FCC--hh. 

A similar situation occurs in the domain of physics detectors, where the planned upgrade to cope with the HL--LHC performance will bring the detectors in a new territory, thus approaching that of the FCC--hh. Hence, also in this domain, FCC--hh can build on the experience of the HL--LHC. 

It is also evident that strong synergy is present between FCC--hh and HL--LHC at the level of beam dynamics, due to the similarity of some regimes. It is possible to identify, as areas with similar challenges, optics control in the ring, in general, and in the experimental insertions, in particular, emittance preservation of high-brightness beams, electron-cloud effects, beam instabilities, as well as, e.g. machine operation with crab cavities.

Finally, it will be easy to find synergies in the domain of energy efficiency and environmental impact with other projects, as these two aspects are gaining so much focus that will become essential items for any large-scale facility for physics research.

\section{Overview of FCC--hh as presented in the 2019 CDR} \label{sec:intro} 

The discovery of the Higgs boson, announced exactly ten years ago, brought to completion the search for the fundamental constituents of matter and interactions that represent the so-called \emph{Standard Model} (SM). Several experimental observations require an extension of the Standard Model. For instance, explanations are needed for the observed abundance of matter over antimatter, the striking evidence for dark matter, and the non-zero neutrino masses. 

Therefore, a novel research infrastructure, based on a highest-energy hadron collider, FCC--hh, with a center-of-mass collision energy of \SI{100}{TeV} and an integrated luminosity of at least a factor of five larger than the HL--LHC~\cite{HL-LHCNature2019,BejarAlonso:2749422} is proposed to address the aforementioned aspects~\cite{FCC-hhCDR,FCC-hhNature2019,FCC-hhNature2020}. The current energy frontier limit for collider experiments will be extended by almost an order of magnitude, and the mass reach for direct discovery will achieve several tens of TeV. Under these conditions, for instance, the production of new particles, whose existence could have emerged from  precision measurements during the preceding e$^+$e$^-$ collider phase (FCC--ee), would become possible. An essential task of this collider will be the accurate measurement of the Higgs self-coupling, as well as the exploration of the dynamics of electroweak symmetry breaking at the TeV scale, to elucidate the nature of the electroweak phase transition. 

This unique particle collider infrastructure, FCC--hh, will serve the world-wide physics community for about $25$~years. However, it is worth stressing that in combination with the lepton collider~\cite{FCC-eeCDR} as initial stage, the FCC integrated project will provide a research tool until the end of the 21st century. 

The FCC construction project will be carried out in close collaboration with national institutes, laboratories, and universities world-wide, with a strong participation of industrial partners. It is worth mentioning that the coordinated preparatory effort is based on a core of an ever-growing consortium of already more than 145 institutes world-wide.

\subsection{Accelerator layout}

The FCC--hh~\cite{FCC-hhCDR,FCC-hhNature2019,FCC-hhNature2020} is designed to provide proton–proton collisions with a center-of-mass energy of \SI{100}{TeV} and an integrated luminosity of $\approx$ \SI{20}{\per \atto \barn} in each of the two primary experiments for $25$~years of operation. The FCC--hh offers a very broad palette of collision types, as it is envisaged to collide ions with protons and ions with ions. The ring design also allows one interaction point to be upgraded to electron–proton and electron–ion collisions. In this case, an additional recirculating, energy-recovery linac will provide the electron beam that collides with one circulating proton or ion beam. The other experiments can operate concurrently with hadron collisions.

The FCC--hh will use the existing CERN accelerator complex as the injector facility. The accelerator chain, consisting of CERN’s Linac4, PS, PSB, SPS, and LHC, could deliver beams at \SI{3.3}{TeV} to the FCC--hh, thanks to transfer lines using \SI{7}{T} superconducting magnets that connect the LHC to FCC--hh. This choice also permits the continuation of CERN’s rich and diverse fixed-target physics programme in parallel with FCC--hh operations. Limited modifications of the LHC should be implemented, in particular, the ramp speed can be increased to optimize the filling time of the FCC--hh. Furthermore, reliability and availability studies have confirmed that operation can be optimized such that the FCC--hh collider can achieve its performance goals. However, the power consumption of the aging LHC cryogenic system is a concern. Note that the required 80\%–90\% availability of the injector chain could best be achieved with a new high-energy booster. As an alternative, direct injection from a new superconducting synchrotron at \SI{1.3}{TeV} that would replace the SPS is also being considered. In this case, simpler normal-conducting transfer lines with magnets operating at \SI{1.8}{T} are sufficient. For this scenario, more studies on beam stability in the collider at injection are required. 

Key parameters of the collider presented in the CDR are given in Table~\ref{tab:param}. In the CDR, the circumference of FCC--hh was \SI{97.75}{km}. Recently a placement optimization has led to a ``lowest-risk'' layout with a circumference of \SI{91.17}{km} (also see Section \ref{sec:progress}
and Fig.~\ref{fig:FCC-hh-current}), comprising four short straight sections of \SI{1.4}{km} length for the experimental insertions, and four longer straight sections of about \SI{2.16}{km} each, that would  house, e.g.~the radiofrequency (RF), collimation, and beam extraction systems.  

Two high-luminosity experiments are located in the opposite insertions PA and PG, which ensures the highest luminosity, reduces unwanted beam-beam effects, and is independent of the beam-filling pattern. The main experiments are located in \SI{66}{m} long experimental caverns, sufficient for the detector that has been studied and ensuring that the final focus system can be integrated into the available length of the insertion. Two additional, lower luminosity experiments are located in the other two experimental insertions.

\begin{table}[htb]
\begin{center}
\caption{Key FCC--hh baseline parameters from the 2019 CDR \cite{FCC-hhCDR} compared to LHC and HL--LHC parameters.}
\begin{tabular}{|l|r|r|r|r|}
\hline 
& \multicolumn{1}{c|}{LHC} & \multicolumn{1}{c|}{HL--LHC} & \multicolumn{2}{c|}{FCC--hh} \\
& & & Initial & Nominal \\
\hline 
\multicolumn{5}{|l|}{Physics performance and beam parameters} \\
\hline
Peak luminosity\tablefootnote{For the nominal parameters, the peak luminosity is reached during the run.} ($10^{34} \SI{}{cm^{-2} s^{-1}}$) & $1.0$ & $5.0$ & $5.0$ & $< 30.0$ \\
Optimum average integrated & $0.47$ & $2.8$ & $2.2$ & $8$ \\
luminosity/day (\SI{}{\femto \barn^{-1}}) & & & & \\
Assumed turnaround time (\SI{}{h}) & & & $5$ & $4$ \\
Target turnaround time (\SI{}{h}) & & & $2$ & $2$ \\
Peak number of inelastic events/crossing & $27$ & $135$\tablefootnote{The baseline assumes leveled luminosity.} & $171$ & $1026$ \\
Total/inelastic cross section & \multicolumn{2}{c|}{$111/85$} & \multicolumn{2}{c|}{$153/108$} \\
$\sigma$ proton (\SI{}{\milli \barn}) & \multicolumn{2}{c|}{} & \multicolumn{2}{c|}{} \\
Luminous region RMS length (\SI{}{cm}) & \multicolumn{2}{c|}{} & \multicolumn{2}{c|}{$5.7$} \\
Distance IP to first quadrupole $L^\ast$ (\SI{}{m}) & \multicolumn{2}{c|}{$23$} & \multicolumn{2}{c|}{$40$} \\
\hline 
\multicolumn{5}{|l|}{Beam parameters} \\
\hline 
Number of bunches $n$ & \multicolumn{2}{c|}{$2808$} & \multicolumn{2}{c|}{$10400$} \\
Bunch spacing (\SI{}{ns}) & \multicolumn{2}{c|}{$25$} & \multicolumn{2}{c|}{$25$} \\
Bunch population $N$ ($10^{11}$) & $1.15$ & $2.2$ & \multicolumn{2}{c|}{1.0} \\
Nominal transverse normalised  & $3.75$ & $2.5$ & \multicolumn{2}{c|}{$2.2$} \\
emittance (\SI{}{\micro m}) & & & \multicolumn{2}{c|}{} \\
Number of IPs contributing to $\Delta Q$ & $3$ & $2$ & $2+2$ & $2$ \\ 
Maximum total beam-beam tune shift $\Delta Q$ & $0.01$ & $0.015$ & $0.011$ & $0.03$ \\
Beam current (\SI{}{A}) & $0.58$ & $1.12$ & \multicolumn{2}{c|}{0.5} \\
RMS bunch length\tablefootnote{The HL--LHC assumes a different longitudinal distribution; the equivalent Gaussian RMS is \SI{9}{cm}.} (\SI{}{cm}) & \multicolumn{2}{c|}{$7.55$} & \multicolumn{2}{c|}{$8$} \\
$\beta^\ast$ (\SI{}{m}) & $0.55$ & $0.15$ & $1.1$ & $0.3$ \\
RMS IP spot size (\SI{}{\micro m}) & $16.7$ & $7.1$ & $6.8$ & $3.5$ \\
Full crossing angle (\SI{}{\micro rad}) & $285$ & $590$ & $104$ & $200$\tablefootnote{The luminosity reduction due to the crossing angle will be compensated using the crab crossing scheme.} \\
\hline
\end{tabular}
\end{center}
\label{tab:param}
\end{table}



The regular lattice in the arc consists of \ang{90} FODO cells with a length of about \SI{213}{m}, six \SI{14}{m}-long dipoles between quadrupoles, and a dipole filling factor of about $0.8$. Therefore, a dipole field around \SI{16}{T} is required to maintain the nominal beams on the circular orbit.

The dipoles are based on Nb$_3$Sn, are operated at a temperature of \SI{2}{K}, and are a key cost item of the collider. Efforts devoted to increasing the current density in the conductors to \SI{1500}{A \per mm^2} at \SI{4.2}{K}, were successful~\cite{uswire1, uswire2}. Several optimized dipole designs have been developed in the framework of the EuroCirCol H2020 EC-funded project. The cosine-theta design has been selected as baseline, because it provided a beneficial reduction of the amount of superconductor needed for the magnet coils. Several collaboration agreements are in place with organisations such as the French CEA, the Italian INFN, the Spanish CIEMAT, and the Swiss PSI, to build short model magnets. It is worth mentioning that a US DOE Magnet Development Programme is actively working to demonstrate a \SI{15}{T} superconducting accelerator magnet and has reached \SI{14.5}{T}.

As the current plans are that FCC--hh is implemented following FCC--ee in the same underground infrastructure, the time scale for design and R\&D for FCC--hh is of the order of 30 years. This additional time will be used to develop alternative technologies, such as magnets based on high-temperature superconductors with a potential significant impact on the collider parameters, relaxed infrastructure requirements (cryogenics system), and increased energy efficiency (temperature of magnets and beam screen).

\subsection{Luminosity performance}

The initial parameters, with a maximum luminosity of $5\times 10^{34}$ \SI{}{cm^{-2} s^{-1}}, are planned to be reached in the first years. Then, a luminosity ramp up will be applied, to reach the nominal parameters with a luminosity of up to $3 \times 10^{35}$ \SI{}{cm^{-2} s^{-1}}. Correspondingly, the integrated luminosity per day will increase from \SI{2}{\femto \barn^{-1}} to \SI{8}{\femto \barn^{-1}}. A luminosity of $2 \times 10^{34}$ \SI{}{cm^{-2} s^{-1}} can be achieved at the two additional experimental insertions, although further studies are needed to confirm this.

High brightness and high-current beams, with a quality comparable to that of the beams of the HL--LHC, combined with a small $\beta^\ast$ at the collision points ensure the high luminosity. The parasitic beam-beam interactions are controlled by introducing a finite crossing angle, whose induced luminosity reduction is compensated by means of crab cavities. Further improvement of the machine performance might be achieved by using electron lenses and current carrying wire compensators.

The fast burn-off under nominal conditions prevents from using the beams for collisions for more than \SI{3.5}{h}. Hence, the turn-around time, i.e. the time from one luminosity run to the next one, is a critical parameter to achieve the target integrated luminosity. In theory, a time of about \SI{2}{h} is within reach, but to include a sufficient margin, turn-around times of \SI{5}{h} and \SI{4}{h} are assumed for initial and nominal parameters, respectively. Note that an availability of 70\% at flat top for physics operation is assumed for the estimate of the overall integrated luminosity.

The collider performance can be affected by various beam dynamics effects that can lead to the development of beam instabilities and quality loss. To fight against these effects a combination of fast transverse feedback and octupoles is used to stabilize the beam against parasitic electromagnetic interaction with the beamline components. Electron cloud build-up, which could render the beam unstable, is suppressed by appropriate hardware design. The impact of main magnet field imperfections on the beam is mitigated by high-quality magnet design and the use of corrector magnets.

\subsection{Technical systems}

Many technical systems and operational concepts for FCC--hh can be scaled up from HL--LHC or can be based on technology demonstrations carried out in the frame of ongoing R\&D projects. Particular technological challenges arise from the higher total energy in the beam (20 times that of LHC), the much increased collision debris in the experiments (40 times that of HL--LHC), and far higher levels of synchrotron radiation in the arcs (200 times that of LHC).

The high luminosity and beam energy will produce collision debris with a power of up to \SI{0.5}{MW} in the main experiments, with a significant fraction of this lost in the ring close to the experiment. A sophisticated shielding system, similar to HL--LHC~\cite{BejarAlonso:2749422}, protects the final focusing triplet, avoids quenches, and reduces the radiation dose. The current radiation limit of \SI{30}{MGy} for the magnets, imposed by the resin used, will be reached for an integrated luminosity of \SI{13}{\atto \barn^{-1}}, but it is projected that the improvement of both the shielding and the radiation hardness of the magnets is possible. Hence, it is likely that the magnets will not have to be replaced during the entire lifetime of the project.

The robust collimation and beam extraction system protects the machine from the energy stored in the beam. The design of the collimation system is based on the LHC system~\cite{LHCDR,BejarAlonso:2749422}, however, with a number of improvements. Additional protection has been added to mitigate losses in the arcs that would otherwise quench magnets. Improved conceptual designs of collimators and dogleg dipoles have been developed to reduce the beam-induced stress to acceptable levels. Further R\&D should aim at gaining margins in the design to reach comfortable levels.

The extraction system uses a segmented, dual-plane dilution kicker system to distribute the bunches in a multi-branch spiral on the absorber block. Novel superconducting septa capable of deflecting the high-energy beams are currently being developed. The system design is fault tolerant, and the most critical failure mode, erratic firing of a single extraction kicker element, has limited impact thanks to the high granularity of the system. Investigations of suitable absorber materials including 3D carbon composites and carbon foams are ongoing in the frame of the HL--LHC project.

The cryogenic system must compensate the continuous heat loads in the arcs of \SI{1.4}{W \per m} at a temperature below \SI{2}{K}, and the \SI{30}{W \per m \per aperture} due to synchrotron radiation at a temperature of \SI{50}{K}, as well as absorbing the transient loads from the magnets ramping. The system must also be able to fill and cool down the cold mass of the machine in less than 20 days, while avoiding thermal gradients higher than \SI{50}{K} in the cryomagnet structure. Furthermore, it must also cope with quenches of the superconducting magnets and be capable of a fast recovery from such situations that leaves the operational availability of the collider at an adequate level. The number of active cryogenic components distributed around the ring is minimized for reasons of simplicity, reliability, and maintenance. Note that current helium cryogenic refrigeration only reaches efficiencies of about 30\% with respect to an ideal Carnot cycle, which leads to high electrical power consumption. For this reason, part of the FCC study is to perform R\&D on novel refrigeration down to \SI{40}{K} based on a neon-helium gas mixture, with the potential to reach efficiencies higher than 40\%, thus bringing a reduction of the electrical energy consumption of the cryogenics system by 20\%.

The cryogenic beam vacuum system ensures excellent vacuum to limit beam-gas scattering, and protect the magnets from the synchrotron radiation of the high-energy beam, also efficiently removing the heat. It also avoids beam instabilities due to parasitic beam-surface interactions and electron cloud effects. Note that the LHC vacuum system design is not suitable for FCC--hh, hence a novel design has been developed in the scope of the EuroCirCol H2020-funded project. It is as compact as possible to minimize the magnet aperture and consequently magnet cost. The beam screen features an anti-chamber and is copper coated to limit the parasitic interaction with the beam; the shape also reduces the seeding of the electron cloud by backscattered photons, and additional carbon coating or laser treatment prevents the build-up. This novel system is operated at \SI{50}{K} and a prototype has been validated experimentally in the KARA synchrotron radiation facility at KIT (Germany).

The RF system is similar to the one of LHC with an RF frequency of \SI{400}{MHz}, although it provides a higher maximum total voltage of \SI{48}{MV}. The current design uses 24 single-cell cavities. To adjust the bunch length in the presence of longitudinal damping by synchrotron radiation, controlled longitudinal emittance blow-up by band-limited RF phase noise is implemented. 

\subsection{Ion operation}

A first parameter set for ion operation has been developed based on the current injector performance. If two experiments operate simultaneously for 30 days, one can expect an integrated luminosity in each of them of \SI{6}{\pico \barn^{-1}} and \SI{18}{\pico \barn^{-1}} for proton-lead ion operation with initial and nominal parameters, respectively. For lead-ion lead-ion operation \SI{23}{\nano \barn^{-1}} and \SI{65}{\nano \barn^{-1}} could be expected, although more detailed studies are in progress to address the key issues in ion production and collimation and to review the luminosity predictions.

\section{Civil engineering}

\begin{figure}[htb]
\begin{center}
\includegraphics[trim=2truemm 0truemm 2truemm 0truemm,width=0.60\hsize,clip=]{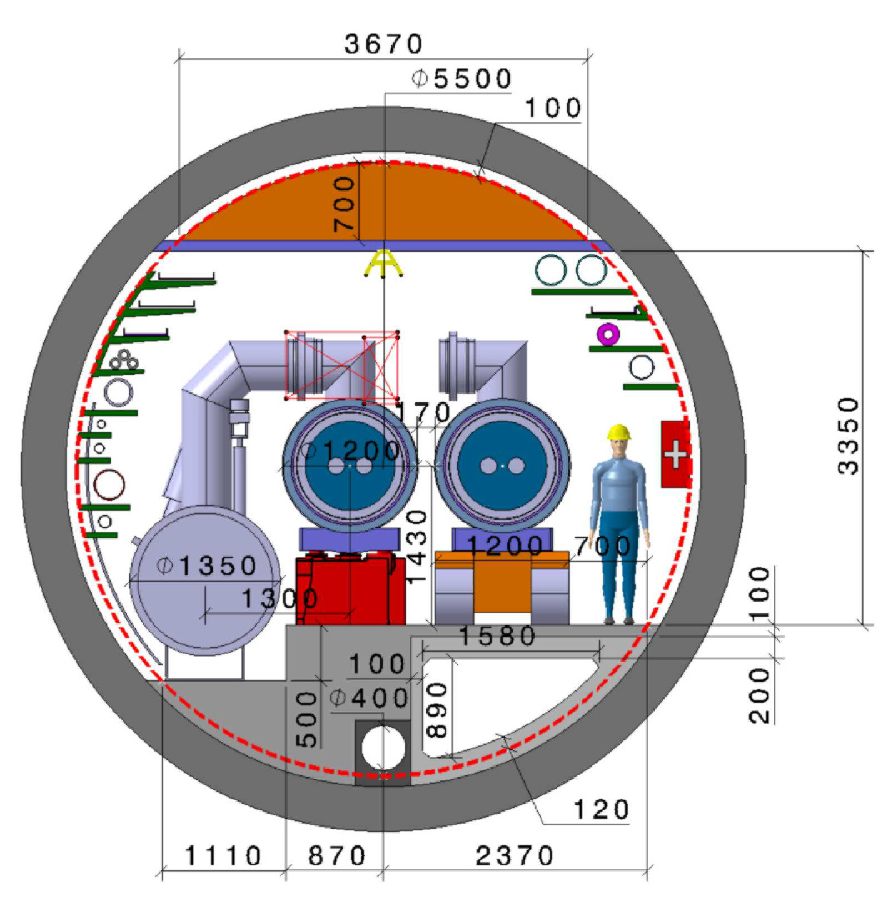}
\end{center}
\caption{Cross section of the FCC--hh tunnel of an arc (from~\cite{FCC-hhCDR}). The gray equipment on the left represents the cryogenic distribution line. A \SI{16}{T} superconducting magnet is shown in the middle, mounted on a red support element. An orange transport vehicle with another superconducting magnet is also shown, in the transport passage.}
\label{fig:tunnel}
\end{figure}

As stated above, the FCC--hh collider will be installed in a quasi-circular tunnel composed of arc segments interleaved with straight sections with an inner diameter of at least \SI{5.5}{m} and a circumference of \SI{91.17}{km}. The internal diameter tunnel is required to house all necessary equipment for the machine, while providing sufficient space for transport and ensuring compatibility between FCC--hh and FCC--ee requirements. Figure~\ref{fig:tunnel} shows the cross section of the tunnel in a typical arc region, including several ancillary systems and services required. Furthermore, about \SI{8}{km} of bypass tunnels, about 18 shafts, 10 large caverns and 8 new surface sites are part of the infrastructure to be built. 

The underground structures should be located as much as possible in the sedimentary rock of the Geneva basin, known as Molasse (which provides good conditions for tunneling) and avoid the limestone of the nearby Jura. Moreover, the depth of the tunnel and shafts should be minimized to control the overburden pressure on the underground structures and to limit the length of service infrastructure. These requirements, along with the constrain imposed by the connection to the existing accelerator chain through new beam transfer lines, led to the clear definition of the study boundary, which should be respected by all possible tunnel layouts considered. A slope of 0.2\% in a single plane will be used for the tunnel to optimize the geology intersected by the tunnel and the depth of the shafts, as well as to implement a gravity drainage system. The majority of the machine tunnel will be constructed using tunnel boring machines, while the sector passing through limestone will be mined. 

The CDR study was based on geological data from previous projects and data available from national services, and based on this knowledge, the civil engineering project is considered feasible, both in terms of technology and project risk control. It is also clear that dedicated ground and site investigations are required during the early stage of the preparatory phase to confirm the findings, to provide a comprehensive technical basis for an optimized placement and as preparation for project planning and implementation processes. It is worth mentioning that for the access points and their associated surface structures, the priority has been given to the identification of possible locations that are feasible from socio-urbanistic and environmental perspectives. Even in this case, the technical feasibility of the construction has been studied and is deemed achievable.

\section{Detector considerations}

The FCC--hh is both a discovery and a precision measurement machine, with the mass reach increased with respect to the current LHC by a factor of seven. The much larger cross sections for SM processes combined with the higher luminosity lead to a significant increase in measurement precision. This implies that the detector must be capable to measure multi-TeV jets, leptons and photons from heavy resonances with masses up to \SI{50}{TeV}, as well as the known SM processes with high precision, and to be sensitive to a broad range of BSM signatures at moderate $p_\mathrm{T}$. Given the low mass of SM particles compared to the \SI{100}{TeV} collision energy, many SM processes feature a significant forward boost while keeping transverse momentum distributions comparable to LHC energies. Hence, a detector for \SI{100}{TeV} must increase the acceptance for precision tracking and calorimetry to $|\eta| \approx 4$, while retaining the $p_\mathrm{T}$ thresholds for triggering and reconstruction at levels close to those of the current LHC detectors. The large number of p–p collisions per bunch crossing, which leads to the so-called pile-up, imposes stringent criteria on the detector design. Indeed, the present LHC detectors cope with pile-up up to 60, the HL--LHC will generate values of up to 200, whereas the expected value of 1000 for the FCC--hh poses a technological challenge. Novel approaches, specifically in the context of high precision timing detectors, will likely allow such numbers to be handled efficiently.

\begin{figure}[htb]
\begin{center}
\includegraphics[trim=2truemm 0truemm 2truemm 0truemm,width=0.90\hsize,clip=]{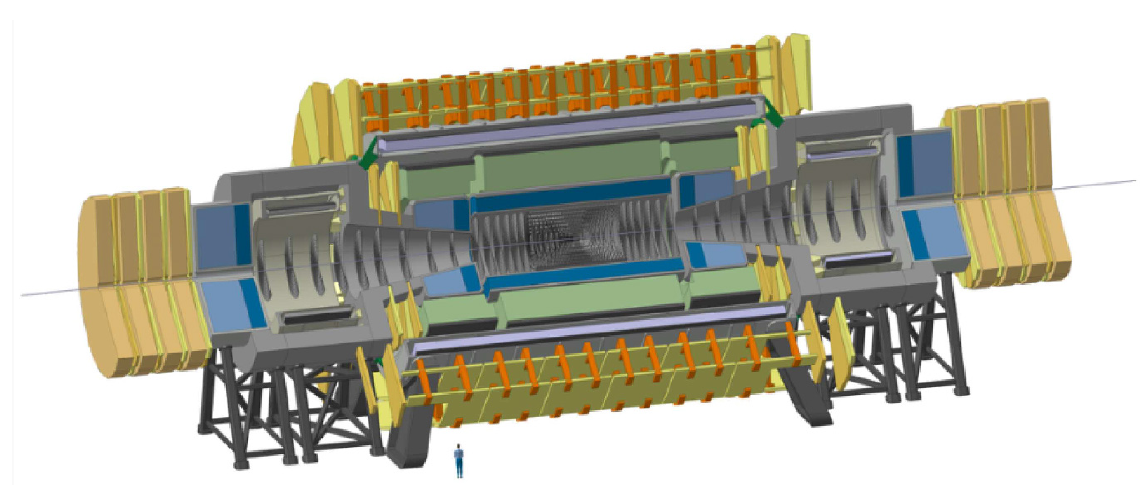}
\end{center}
\caption{Conceptual layout of the FCC--hh reference detector (from~\cite{FCC-hhCDR}). It features an overall length of \SI{50}{m} and a diameter of \SI{20}{m}. A central solenoid with \SI{10}{m} diameter bore and two forward solenoids with \SI{5}{m} diameter bore provide a \SI{4}{T} field for momentum spectroscopy in the entire tracking volume.}
\label{fig:detector}
\end{figure}

Figure~\ref{fig:detector} shows the conceptual FCC--hh reference detector, which serves as a concrete example for subsystem and physics studies aimed at identifying areas where dedicated R\&D efforts are needed. The detector has a diameter of \SI{20}{m} and a length of \SI{50}{m}, similar to the dimensions of the ATLAS detector at the LHC. The central detector, with coverage of $|\eta| < 2.5$, houses the tracking, electromagnetic calorimetry, and hadron calorimetry surrounded by a \SI{4}{T} solenoid with a bore diameter of \SI{10}{m}. The required performance for $|\eta| > 2.5$ is achieved by displacing the forward parts of the detector away from the interaction point, along the beam axis. Two forward magnet coils, generating a \SI{4}{T} solenoid field, with an inner bore of \SI{5}{m} provide the required bending power. Within the volume covered by the solenoids, high-precision momentum spectroscopy up to $|\eta| \approx 4$ and tracking up to $|\eta| \approx 6$ is ensured. Alternative layouts concerning the magnets of the forward region are also studied~\cite{FCC-hhCDR}.

The tracker is specified to provide better than 20\% momentum resolution for $p_\mathrm{T}=\SI{10}{TeV/c}$ for heavy $Z'$ type particles, and better than 0.5\% momentum resolution at the multiple scattering limit, at least up to $|\eta| = 3$. The tracker cavity has a radius of \SI{1.7}{m} with the outermost layer at around \SI{1.6}{m} from the beam, providing the full spectrometer arm up to $|\eta| = 3$. The electromagnetic calorimeter (EMCAL) uses a thickness of around 30 radiation lengths, and together with the hadron calorimeter (HCAL), provides an overall calorimeter thickness of more than 10.5 nuclear interaction lengths to ensure 98\% containment of high-energy showers and to limit punch-through to the muon system.

The EMCAL is based on liquid argon (LAr) due to its intrinsic radiation hardness.
The barrel HCAL is a scintillating tile calorimeter with steel and Pb absorbers,
divided into a central and two extended barrels. The HCALs for the endcap
and forward regions are also based on LAr. The requirement of calorimetry acceptance
up to $|\eta| \approx 6$ translates into an inner active radius of only \SI{8}{cm} at a $z$-distance of \SI{16.6}{m} from the interaction point. The EMCAL is specified to have an energy resolution around $10\%/\sqrt{E}$, while the HCAL around $50\%/\sqrt{E}$ for single particles. The features of the muon system have a significant impact on the overall detector design. As nowadays there is little doubt that large-scale silicon trackers will be core parts of future detectors, the emphasis on standalone muon performance is less pronounced, and the focus is rather shifted towards the aspects of muon trigger and muon identification.

In the reference detector, the magnetic field is unshielded, with several positive side effects that concur to a sensible cost reduction. The unshielded coil can be lowered through a shaft of \SI{15}{m} diameter and the detector can be installed in a cavern of \SI{37}{m} height and \SI{35}{m} width, similar to the present ATLAS cavern. The magnetic stray field reaches \SI{5}{mT} at a radial distance of \SI{50}{m} from the beamline, so that no relevant stray field leaks in the service cavern, placed \SI{50}{m} away from the experiment cavern and separated by rock. The shower and absorption processes inside the forward calorimeter produce a large number of low-energy neutrons, a significant fraction of which enters the tracker volume. To keep these neutrons from entering the muon system and the detector cavern, a heavy radiation shield is placed around the forward solenoid magnets to close the gap between the endcap and forward calorimeters.

The technologies selected for the various subsystems should stand significant radiation levels. On the first silicon layer at $r = \SI{2.5}{cm}$ the charged-particle rate is around \SI{10}{GHz \per cm^2}, and it drops to about \SI{3}{MHz \per cm^2} at the outer radius of the tracker, whereas inside the forward EMCAL the number rises to \SI{100}{GHz \per cm^2}. The \SI{1}{MeV} neutron equivalent fluence, a key number for long-term damage of silicon sensors and electronics in general, evaluates to a value of $6 \times 10^{17} \SI{}{\per cm^2}$ for the first silicon layer, beyond a $ r = \SI{40}{cm}$ the number drops below $10^{16}\SI{}{\per cm^2}$, and in the outer parts of the tracker it is around $5 \times 10^{15} \SI{}{\per cm^2}$. This means that technologies used for the HL--LHC detectors are therefore applicable when $r > \SI{40}{cm}$, while novel sensors and readout electronics have to be developed for the innermost parts of the tracker.

The charged particle rate in the muon system is dominated by electrons, created from high energy photons in the \SI{}{MeV} range by processes related to thermalization and capture of neutrons that are produced in hadron showers mainly in the forward region. In the barrel muon system and the outer endcap muon system, the charged particle rate does not exceed \SI{500}{Hz \per cm^2}, the rate in the inner endcap muon system increases to \SI{10}{kHz \per cm^2}, and to \SI{500}{kHz \per cm^2} in the forward muon system, at a distance of \SI{1}{m} from the beam. These rates are comparable to those of the muon systems of the current LHC detectors, therefore, gaseous detectors used in these experiments can be adopted.

\section{Cost and schedule}

In the FCC integrated project, the FCC--hh is preceded by the lepton collider Higgs, top, and electroweak factory, FCC--ee. Here,  both civil engineering and general technical infrastructures of the FCC--ee can be fully reused for FCC--hh, thus substantially lowering the investments for the latter to \SI{17000}{MCHF}, according to the CDR estimate \cite{FCC-hhCDR}.  The particle collider- and injector-related investments amount to 80\% of the FCC--hh cost, namely to about \SI{13600}{MCHF}. The major part of this accelerator cost corresponds to the expected price of the 4700 Nb$_3$Sn \SI{16}{T} main dipole magnets, totaling \SI{9400}{MCHF}, for a target cost of \SI{2}{MCHF}/magnet. For completeness, we note that in the CDR, the construction cost for FCC--hh as a single standalone project, i.e.\  without prior construction of an FCC--ee lepton collider, was estimated to be about \SI{24000}{MCHF} for the entire project.

The FCC--hh operation costs, other than electricity cost, are expected to remain limited, based on the evolution from LEP to LHC operation today, which shows a steady decrease in the effort needed to operate, maintain and repair the equipment. The cost-benefit analysis of the LHC/HL--LHC programme reveals that a research infrastructure project of such a scale and high-tech level has the potential to generate significant socio-economic value throughout its lifetime, in particular if the tunnel, surface, and technical infrastructures from a preceding project have been amortized.

In the integrated FCC project, disassembly of the FCC--ee and subsequent installation of the FCC--hh take about 8--10 years. The projected duration for the operation of the FCC--hh facility is 25 years, to complete the currently envisaged proton-proton collision physics programme. As a combined, ``integrated'' project, namely FCC--ee followed by FCC--hh, the FCC covers a total span of at least 70 years, i.e.~until the end of the 21st century.

\section{Progress since the CDR} \label{sec:progress}

\subsection{Evolution of the baseline layout}

\begin{figure}[htb]
\begin{center}
\includegraphics[trim=50truemm 0truemm 50truemm 0truemm,width=0.80\hsize,clip=]{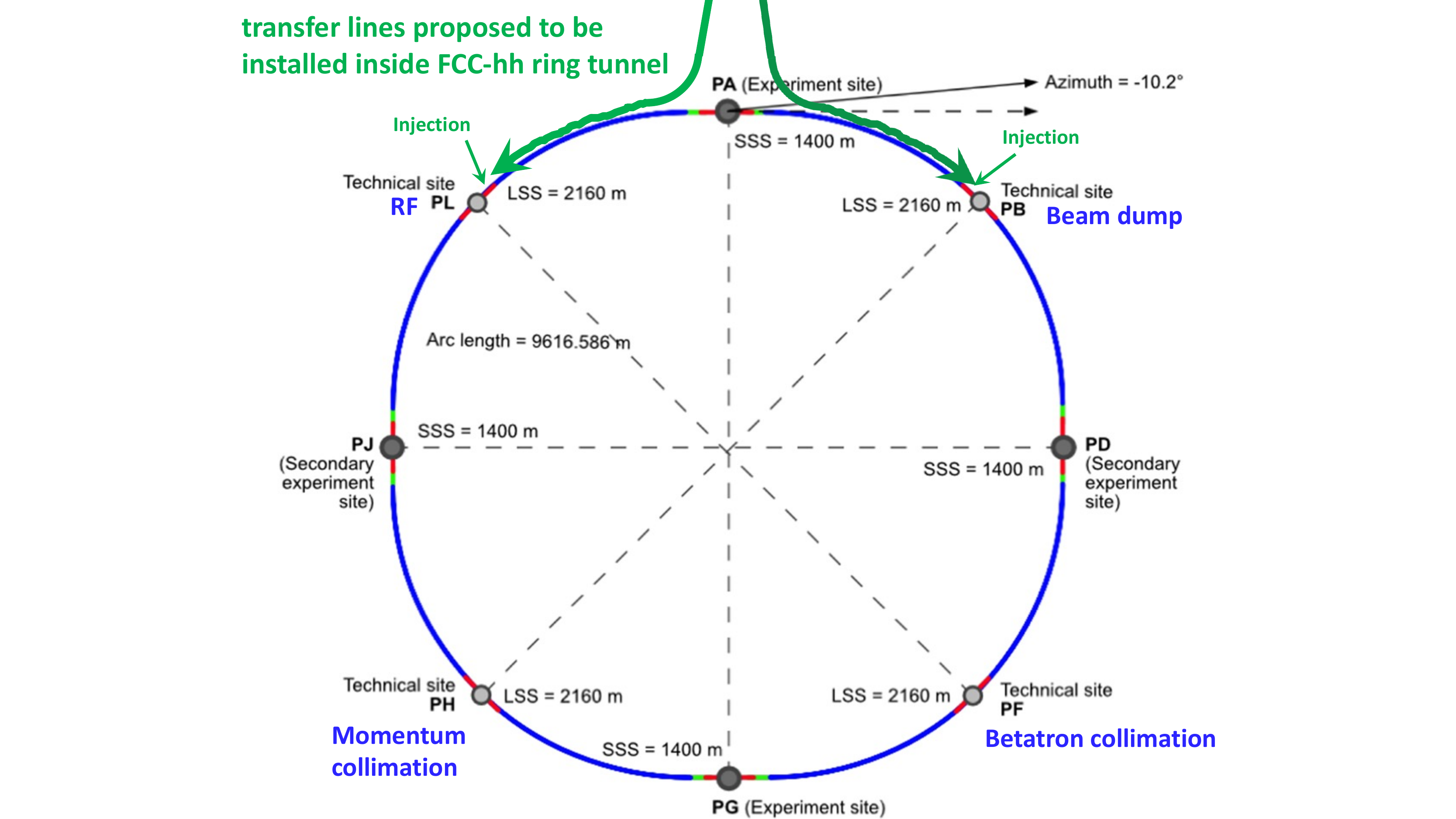}
\end{center}
\caption{Sketch of the proposed eight-point FCC--hh layout.}
\label{fig:FCC-hh-current}
\end{figure}

Among the several domains of activity that have been pursued since the publication of the CDR, it is important stressing the intense efforts devoted to placement studies, which refined the results discussed in~\cite{FCC-hhCDR}. These aim to determine an optimal tunnel layout that could fulfill the multiple constraints imposed by geological situation, territorial and environmental aspects. Furthermore, in the frame of FCC--ee studies, it emerged that implementing four experimental interaction points is an interesting option worth investigating. Beam dynamics considerations impose a symmetrical positioning of the four experimental points. Hence, to allow sharing the experimental caverns between FCC--ee and its hadron companion, the same principle should also be applied to the FCC--hh lattice. The outcome of these considerations is the new layout shown in Fig.~\ref{fig:FCC-hh-current}.  The circumference of the proposed layout is \SI{91.17}{km}. The proposed layout has an appealing side effect, namely, only eight access points are present, with a non-negligible impact on the civil engineering works and costs.

The four experimental points are located in PA, PD, PG, and PJ, respectively. The length of the straight sections has been revised, following the results of the placement studies: a short straight section, \SI{1.4}{km} in length like in the baseline lattice, is used to house the experimental interaction points; a long straight section, \SI{2.16}{km} in length, is used to house the key systems. Currently, it is proposed to install the beam dump in PB, the betatron collimation in PF, the momentum collimation in PH, and the RF system in PL. These preliminary assignments should be confirmed by detailed studies. Such studies should also assess the feasibility of the optics required for the various systems, following the sizable length reduction (from \SI{2.8}{km} of the CDR baseline version to \SI{2.16}{km} for the new version). 

The total length of the arcs is \SI{76.93}{km}, and, unlike the baseline configuration, all arcs have the same length. The reduction of the total arc length implies that the collision energy falls short of \SI{100}{TeV} by few TeV, and this is not felt as a hurdle. The FODO cell length is unchanged.  

The rearrangement of the experimental points has an impact on the injection and transfer line design. The configuration inherited from the LHC design, in which the injection is performed in the same straight section in which the secondary experiments are installed, has to be dropped as it would lead to very long transfer lines. Therefore, the current view consists of combining the injection with beam dump (in PB) and with RF (in PL). Then, to save in tunnel length, it is proposed that the transfer lines run in the FCC--hh ring tunnel from close to PA until the injection point (see Fig.~\ref{fig:FCC-hh-current}). An additional benefit of this solution is that the transfer line magnets would be normal-conducting and rather relaxed in terms of magnetic properties. Integration of the transfer lines in the ring tunnel is being actively pursued to assess the feasibility of this proposal.  

\subsection{Alternative configuration}

In parallel to the studies for the optimization of the baseline layout, some efforts have been devoted to the analysis of alternative approaches to the generation of the ring optics. Indeed, the standard paradigm to collider optics consists in using separate-function magnets, in particular in the regular arcs, in conjunction with a FODO structure. However, a combined-function optics might provide interesting features, particularly appealing for an energy-frontier collider. A combined-function optics has the potential of providing a higher dipole filling factor, thus opening to interesting optimization paths of the dipole field and beam energy. Currently, this research has explored the benefits of a combined-function periodic cell~\cite{our_paper6}. It also optimized some of the parameters of the cell, such as its length~\cite{our_paper8}, showing that the combined-function magnet is equally feasible as the baseline magnet. Furthermore, a complex optics, including arc and dispersion suppressors, can indeed be realized with combined-function magnets. As a next step, the investigations will consider the various systems of corrector magnets planned in the baseline FODO cell and optimize them in the context of a combined-function periodic cell. 

\section{Conclusions}

The FCC--hh baseline comprises a power-saving, low-temperature superconducting magnet system based on an evolution of the Nb$_3$Sn technology pioneered at the HL--LHC. An energy-efficient cryogenic refrigeration infrastructure, based on a neon-helium light gas mixture, and a high-reliability and low-loss cryogenic distribution infrastructure are also key elements of the baseline. Highly segmented kickers, superconducting septa and transfer lines, and local magnet energy recovery, are other essential components of the proposed FCC--hh design. Furthermore, technologies that are already being gradually introduced at other CERN accelerators will be deployed in the FCC--hh. Given the time scale of the FCC integrated program that allows for around 30 years of R\&D for FCC-hh, an increase of the energy efficiency of a particle collider can be expected thanks to high-temperature superconductor R\&D, carried out in close collaboration with industrial partners. The reuse of the entire CERN accelerator chain, serving also a concurrent physics programme, is an essential lever to come to an overall sustainable research infrastructure at the energy frontier.

The FCC--hh will be a strong motor of economic and societal development in all participating nations, because of its large-scale and intrinsic character of international fundamental research infrastructure, combined with tight involvement of industrial partners. Finally, it is worth stressing the training provided at all education levels by this marvelous scientific tool. 





\bibliographystyle{JHEP}
\bibliography{mybibliography}  


\end{document}